\documentclass[sigconf, nonacm]{acmart}
\renewcommand\footnotetextcopyrightpermission[1]{}  
\usepackage{amsmath,amsfonts}
\usepackage{algorithmic}
\usepackage{graphicx}
\usepackage{textcomp}
\usepackage{xcolor}
\usepackage{xspace}
\usepackage{pifont}
\usepackage{tikz}
\usepackage{algorithm}
\usepackage{algorithmic}
\def\BibTeX{{\rm B\kern-.05em{\sc i\kern-.025em b}\kern-.08em
    T\kern-.1667em\lower.7ex\hbox{E}\kern-.125emX}}

\begin{document}

\newcommand{\circnum}[1]{\tikz[baseline=(char.base)]{
  \node[shape=circle,draw,inner sep=0.4pt,font=\small, fill=black, text=white] (char) {#1};}}

\newcommand*\circled[1]{\tikz[baseline=(c.base)]{%
  \node[shape=circle, fill=black, text=white,
        inner sep=0.8pt, minimum size=1.4ex,
        font=\scriptsize\bfseries] (c) {#1};}}
\title{XMPIaaS: Towards Cloud Native MPI via Cooperative Process Migration}

\author{Shunyu Yao}
\email{shunyu@vt.edu}
\affiliation{%
\institution{Virginia Tech}
  \city{Blacksburg}
  \state{VA}
  \country{USA}
}

\author{Dimitrios S. Nikolopoulos}
\email{dsn@vt.edu}
\affiliation{%
\institution{Virginia Tech}
  \city{Blacksburg}
  \state{VA}
  \country{USA}
}

\author{Ali R. Butt}
\email{butta@vt.edu}
\affiliation{%
\institution{Virginia Tech}
  \city{Blacksburg}
  \state{VA}
  \country{USA}
}







\newcommand{\name}{XMPIaaS\xspace}
\begin{abstract}
Message Passing Interface (MPI) has been the dominant programming model for High Performance Computing (HPC) for three decades, and as HPC workloads increasingly migrate to cloud infrastructure for scalability and cost efficiency, MPI applications must contend with an execution environment fundamentally unlike traditional supercomputers: ephemeral resources, dynamic pricing and preemptable instances. In such a volatile setting, the ability to relocate running MPI processes between nodes without restarting the job is a necessity for cost-effective, resilient execution. Existing approaches either require restarting the entire job from a global checkpoint, or transparently intercepting the full MPI stack at prohibitive complexity. To address these challenges, we propose \name, a cooperative migration system for MPI that enables selective process group migration on-the-fly. When a cloud instance is scheduled for preemption, only the affected ranks are relocated while the remaining processes briefly quiesce and resume in place, avoiding the cost of a full-job checkpoint. \name tackles this through a cooperative protocol between the MPI process management runtime and rank processes. We expose an \texttt{XMPI\_quiesce} interface built atop the MPI Sessions API that allows applications to mark safe migration points, and we extend the Hydra process manager to orchestrate the full migration lifecycle: rank quiescence, CRIU checkpoint/restore, proxy relaunch on the target node, and seamless rank reconnection. We evaluate and show that the cooperative quiesce phase accounts for less than
1.4\% of total migration downtime, and that this downtime is governed by the migrating node's rank count alone, independent of job size, and the instrumentation introduces no measurable overhead
during normal execution.

\end{abstract}

\ccsdesc[500]{Computer systems organization~Cloud computing}
\ccsdesc[500]{Computing methodologies~Parallel programming languages}
\ccsdesc[500]{Computer systems organization~Reliability}

\maketitle

\section{Introduction}
\label{sec:intro}

The Message Passing Interface (MPI) has served as the de-facto standard \cite{mpi40, gropp1994using, gropp1996mpich, bernholdt2020survey, bernholdt2024ompix} for parallel communication in High Performance Computing (HPC), forming the backbone of large-scale scientific simulations. Traditionally, MPI applications are deployed on dedicated supercomputers and HPC clusters under a rigid model with static resource allocation and exclusive node  \cite{yoo2003slurm}. As cloud platforms evolve to offer increasingly powerful compute, HPC communities are increasingly migrating their workloads to cloud platforms being attracted by on-demand scalability, resource elasticity and flexible pricing \cite{netto2018hpc, voorsluys2012spot}. 

However, the conventional deployment model of MPI applications is fundamentally at odds with the cloud computing paradigm, where resources are ephemeral and preemption can happen at any time. Several cloud deployment models exist for HPC workloads, each presenting distinct challenges for MPI. The simplest way of renting dedicated virtual machines or bare metals in an Infrastructure-as-a-Service (IaaS) model offers little advantage over traditional clusters beyond hardware procurement convenience, resources remain statically allocated and exclusively held \cite{netto2018hpc}. Preemptible instances (AWS Spot, Azure Spot) reduce costs for up to 90\% \cite{awsspot, azurespot}, but the provider may reclaim any node at any time, and MPI's tightly coupled runtime and processes cannot survive partial node loss. Serverless and Function-as-a-Service platforms \cite{spillner2018faaster, jonas2019cloud} offer the most cloud-native execution model with automatic scaling, per-invocation billing, and zero infrastructure management, but it imposes strict resource limits per instance and assume stateless, short-lived workloads, fundamentally incompatible with MPI's long-running, stateful processes \cite{hellerstein2019serverless}. Across all these models, the common barrier is the same: MPI lacks a mechanism for individual processes to survive the preemption and resource boundaries in a volatile cloud environment. The natural solution is to evacuate affected processes from terminating instances and migrate them onto available ones for continued execution.

Enabling migration for MPI ranks — the individual MPI worker processes — is difficult for three reasons. First, MPI ranks maintain complex network states from TCP socket buffers to RDMA queue pairs, memory registrations, and completion queues. These states are unfit for checkpointing because they are opaque to userspace \cite{criu, planeta2021migros}. General-purpose checkpoint tools cannot serialize these states, and relocating a rank to a new node renders them stale. Thus, network states must be torn down at checkpoint and reconstructed at restore.
Second, MPI ranks are tightly coupled. Different MPI applications have their distinct rank communication patterns, and MPI ranks synchronize through collective operations. Therefore, migrating ranks breaks the collectively established communication topology, disrupts applications' embedded synchronization pattern, and requires coordinated action across the entire job to participate in rebuilding new connections. 
Third, existing MPI process managements are designed around a static process-to-node mapping established at launch time \cite{hydra, castain2018pmix}, they provide no mechanism for a rank to change its hosting node, update its network identity, or rejoin the MPI process group from a new location mid-execution.

Prior works have sought to address resilience for MPI applications\cite{bland2013ulfm, losada2020ulfm, laguna2016reinit, moody2010scr, bautista2011fti, nicolae2019veloc, garg2019mana, xu2023mana2, ansel2009dmtcp, hargrove2006blcr, ropars2013spbc, chakravorty2006proactive, wang2008proactive, wang2025livemigration} (See Section \ref{sec:related_work}). Representatively,
MANA \cite{garg2019mana} achieves transparent checkpoint/restore by virtualizing internal MPI states, using a split-process architecture to isolate the MPI library and discard network context at checkpoint time. The transparency costs tracking and replaying every MPI object's lifecycle, and requires the entire job to be checkpointed no matter the granularity of disruption.
Charm++ and AMPI \cite{kale1993charm, kale1996charm, huang2003ampi, chakravorty2006proactive, bhosale2025charm, bhosale2025elastic, gupta2025charm} decompose a job into migratable virtual ranks that the runtime's load balancer can relocate, evacuating a node without changing the rank count the application sees. However, AMPI requires privatizing global and static variables and an over-decomposed launch, and it relocates its own virtual ranks rather than a stock MPI process.
User-Level Fault Mitigation (ULFM) \cite{bland2013ulfm} extends the MPI standard with fault-detection and communicator-repair semantics, allowing applications to survive process failures. But ULFM provides no mechanism to preserve or relocate the ranks' progress, and it requires applications to explicitly implement its own recovery logic from scratch. 
The ability to evacuate only affected instances and minimize disruption to the rest of the job for modern MPI is left unaddressed.

We propose \name, a cooperative framework for MPI that enables the selective migration of the MPI process group at the execution time. At the time of preemption, \name allows provider to issue notifications to MPI process management proxies and MPI ranks on only affected instances, thus avoiding the cost of a full-job checkpoint. Then \name orchestrates a migration protocol between MPI application processes and the MPI process manager. On the application side, \name exposes an \texttt{XMPI\_quiesce} interface built atop the MPI Sessions API \cite{mpi40}, allowing developers to mark safe migration points at natural iteration boundaries. When triggered, all ranks collectively tear down their network resources and park at the quiesce point. After the affected ranks have been relocated, all ranks synchronize and jointly open a fresh session before resuming computation. Ranks that are not migrating never undergo checkpointing, they simply wait at the quiesce point and rejoin once the relocated ranks have been restored. On the process manager side, we extend MPI's process management layer to coordinate the migration: spinning up a replacement proxy on the destination node, snapshotting only the affected ranks using CRIU \cite{criu}, performing an atomic control-channel handover from the departing proxy to its replacement, and restoring each rank with substituted process management control sockets. We implement \name on top of MPICH's Hydra process manager \cite{hydra, gropp1996mpich}, though its design relies only on abstractions common to all major MPI process managers and is portable to other implementations such as Open MPI's PRRTE \cite{castain2018pmix}.

Three key insights underpin \name's design. First, at a cooperative quiesce point, the process is reduced to its simplest possible form: application memory and a single process management socket with no network connections. This makes all the rank states local and trivially checkpointable by CRIU without any MPI-specific plugins or virtualization. Second, migration is per-instance, not per-job: only the ranks on the affected node are checkpointed and relocated, while non-migrating ranks experience a brief quiesce and resume in place, avoiding the cost of a full-fleet checkpoint. Third, we build \name on top of MPI libraries, with design that is easily portable to other standard-conforming MPI library and process manager implementations. Together, these properties make \name a natural fit for the cloud, where preemption strikes individual instances with little warning.

The main contributions of this paper are summarized as follows:
\begin{itemize}
\item We design an application-side cooperative quiesce built on the MPI Sessions API that allows all ranks to tear down and reconstruct their MPI state within a single process lifetime.
\item We design and implement a runtime-side migration orchestration protocol within MPICH's Hydra process manager that coordinates rank quiescence, selective checkpointing, proxy replacement, and seamless process group reconstruction.
\item We evaluate \name on a multi-node cluster with four proxy applications
and show that the cooperative quiesce phase accounts for less than
1.4\% of total migration downtime, image transfer dominates at
60--92\%, and migration downtime is governed by the migrating node's rank count alone, independent of job size, the instrumentation introduces no measurable overhead
during normal execution.
\end{itemize}                                                                                                                             

\section{Background}
\label{sec:background}

\subsection{Cloud Deployment Models for HPC}
\label{sec:cloud}

Cloud platforms offer several deployment models for compute-intensive workloads, each presenting different tradeoffs for MPI applications. The most straightforward is dedicated Infrastructure-as-a-Service (IaaS), where users rent virtual machines or bare-metal instances for the duration of a job \cite{netto2018hpc}. This model mirrors traditional cluster computing and is fully compatible with MPI, but resources remain statically allocated and exclusively held.

Preemptible instances (Spot Instances on AWS, Spot VMs on Azure, and Spot VMs on Google Cloud) provide access to spare cloud capacity at substantially reduced cost \cite{awsspot, azurespot, gcpspot}. Amazon EC2 Spot Instances offer up to 90\% discount compared to on-demand prices \cite{awsspot}, and Azure Spot VMs similarly offer discounts of up to 90\% off standard pay-as-you-go prices \cite{azurespot}. The tradeoff is that when Azure needs capacity back, the infrastructure will evict Spot VMs with 30 seconds notice \cite{azurespot}, while AWS provides a two-minute warning before reclaiming a Spot Instance \cite{awsspot, awshpcspot}. This preemption model is designed for stateless, fault-tolerant workloads, it is not practical to run parallel MPI jobs or stateful HPC services on these low-cost instances under current tooling \cite{he2012spotmpi, voorsluys2012spot}. MPI's tightly coupled processes cannot survive partial node loss. When one instance is reclaimed, all ranks holding pairwise connections to the evicted ranks are left with broken links, and the entire job typically must be restarted. Enabling MPI to tolerate preemption would unlock a significant portion of cloud capacity for scientific computing.

Serverless and Function-as-a-Service (FaaS) platforms represent the most cloud-native execution model, offering automatic scaling, per-invocation billing, and zero infrastructure management \cite{awslambda, azurefunctions, gcpfunctions, jonas2019cloud}. Today, serverless computing poses challenges for HPC workloads due to resource limits imposed by cloud providers, including maximum memory, CPU, and runtime restrictions \cite{spillner2018faaster, copik2023rfaas}. More fundamentally, FaaS platforms contracts stateless, short-lived function invocations, while MPI requires long-running processes with persistent state and direct inter-process communication \cite{hellerstein2019serverless}. A long-running MPI simulation cannot complete within a single function invocation's lifetime. However, a migration-capable MPI runtime could bridge this gap and survive the resource boundaries of serverless instances. If ranks can be checkpointed before a function's time limit expires and restored into a fresh invocation, the serverless platform effectively becomes a source of renewable short-lived compute slots stitched together into a long-running job \cite{miao2023spotserve}. This reframing transforms serverless from an incompatible execution model into a viable, elastically scaled substrate for MPI, provided that migration overhead remains small relative to the function's available runtime.

\subsection{MPI Fault Tolerance}
\label{sec:mpi_fault_tolerance}

Prior work on MPI resilience imposes its cost in one of three places. \textbf{Transparent checkpoint/restart} places the cost on the MPI implementation, preserving a rank across a checkpoint requires interposing on the MPI interface, virtualizing and replaying internal MPI state, and supplying transport-specific machinery to discard and rebuild network contexts that cannot be serialized \cite{garg2019mana, xu2023mana2, ansel2009dmtcp, huang2003ampi}. This requires either MPI implementation dependence or explicit application rewrite against a dedicated customized execution model. \textbf{Malleability and communicator-repair} approaches instead place the cost on the application requiring user to modify compute logic to tolerate rank shrinking and expansion, redistributing data into a new decomposition, or supplying explicit recovery logic \cite{compres2016elastic, iserte2018dmr, iserte2021dmrlib, huber2024dpp, bland2013ulfm, losada2020ulfm}. \textbf{Whole-job checkpoint/restart} has a cascading cost due to that a preemption confined to limited instance can trigger a checkpoint, a rollback, or a reconfiguration across every rank, at an expense that scales with the size of the job rather than the size of the loss \cite{moody2010scr, bautista2011fti, garg2019mana, ropars2013spbc}. There still lacks an approach with balanced tradeoff among the above cost to accommodate MPI in the volatile environment on cloud. 

\subsection{MPI Sessions}
\label{sec:mpi_sessions}
The MPI 4.0 standard \cite{mpi40} introduced the Sessions model as an alternative to the traditional MPI\_Init/MPI\_Finalize lifecycle. Under the legacy model, an application initializes MPI exactly once and finalizes it exactly once, there is no standard mechanism to tear down MPI state and reinitialize it within the same process for multiple times. With the Sessions API, MPI resource management is decoupled from process lifetime, and an application may repeat arbitrary rounds of Session creation (MPI\_Session\_init), obtaining an MPI process group from a named process set (MPI\_Group\_from\_session\_pset), constructing a communicator (MPI\_Comm\_create\_from\_group), and finalizing the session (MPI\_Session\_finalize). Each session cycle yields a fresh set of MPI resources with no residual state from the previous session, and is effectively equivalent to a complete MPI\_Init/MPI\_Finalize lifecycle. This ability to repeatedly tear down and reconstruct MPI state within a single process lifetime opens door for \name's migration scheme that must preserve a rank's application state across relocation without restarting the process.

\subsection{MPI Process Management}
\label{sec:mpi_process_management}
MPI implementations rely on a process management layer to launch, monitor, and coordinate rank processes throughout a job's lifetime. In MPICH, this role is filled by Hydra \cite{hydra}, which follows a three-tier architecture. A central launcher process (mpiexec) spawns one proxy (pmip) per node on each allocated node via SSH or a resource manager such as SLURM. Each proxy in turn forks and executes the rank processes locally on its node. The internal mechanisms can be explained at Section \ref{sec:pm_hydra}.

Ranks communicate with this process management hierarchy through the Process Management Interface (PMI) \cite{balaji2010pmi, castain2018pmix}, a wire protocol between MPI rank processes and their process manager. Through PMI, ranks perform key-value store operations, fence barriers, and process lifecycle commands such as spawn. Its primary role during initialization is coordinating the exchange of endpoint addresses needed for communicator construction. The per-node proxy acts as a relay, forwarding PMI requests upstream to the central launcher, which serves as the authoritative key-value store and barrier coordinator. PMI has evolved through three generations: PMI-1, PMI-2, and PMIx \cite{balaji2010pmi, castain2018pmix}. They differ in framing, scalability optimizations, and feature set, but all three share the same core semantics.

\subsection{CRIU}
\label{sec:criu}
CRIU \cite{criu} is a Linux tool that can freeze a running process, serialize its states of memory pages, file descriptors, register contents, signal masks, and memory mappings to a set of image files on disk, and later restore the process from those images, potentially on a different node. CRIU operates mainly in userspace, leveraging kernel interfaces such as ptrace, /proc, and prctl to capture and reconstruct process state. CRIU can checkpoint standard POSIX resources such as regular files, pipes, Unix sockets, TCP connections. CRIU also provides a file-descriptor inheritance mechanism (criu\_add\_inherit\_fd) that allows the restoring process to receive fresh file descriptors in place of the originals. 

However, CRIU cannot serialize state that lives in kernel subsystems or hardware devices without explicit plugin support. Notably, CRIU does not support InfiniBand \cite{planeta2021migros}. Device file descriptors, kernel-side RDMA objects (queue pairs, completion queues, memory registrations), and device-backed memory mappings are all opaque to CRIU's checkpoint machinery.

\section{Process Management in Hydra}
\label{sec:pm_hydra}

This section describes the internal architecture of MPICH's Hydra process manager that \name repurposes for migration. We introduce the three-tier topology of process management launcher, process management proxy, and application rank by walking through the process management lifecycle of startup (\ref{sec:startup1} and \ref{sec:startup2}) and teardown (\ref{sec:teardown}).
                                                                     
\subsection{Architecture}
\label{sec:startup1}

\begin{figure}[t]
\centering
\includegraphics[width=\columnwidth]{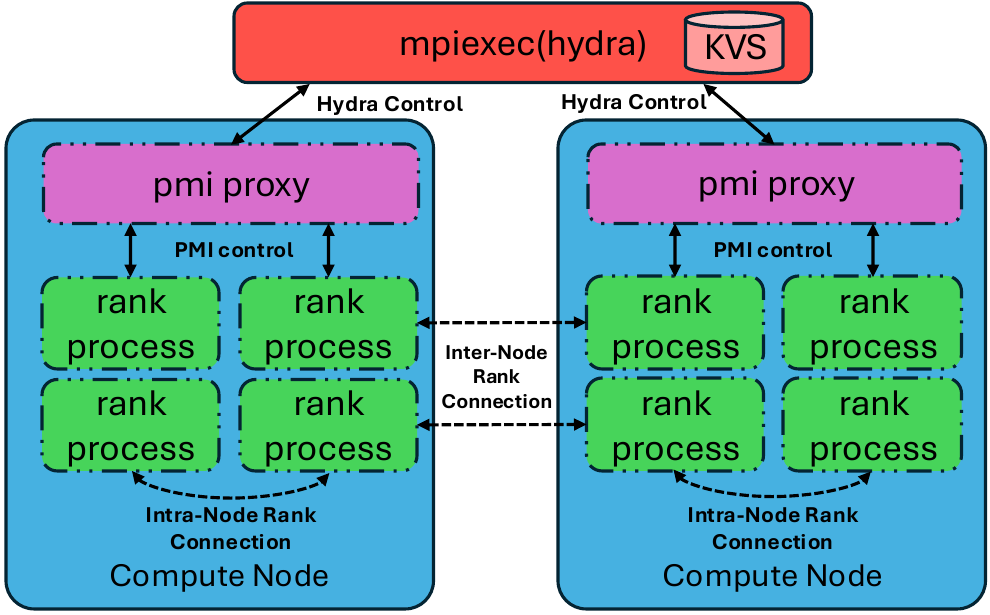}
\caption{Hydra's three-tier process management topology. The TCP control channel connects mpiexec to each pmip, while a Unix socketpair carries PMI messages between each pmip and its local ranks.}
\label{fig:hydra_topology}
\end{figure}

Hydra's central design choice is a strict separation between process management and computation: all coordination state resides in a single launcher and the per-node proxies, while ranks hold no management state of their own. This separation arises from a three-tier hierarchy shown in Figure \ref{fig:hydra_topology}. launcher (mpiexec) sits at the top as the job launcher and central coordinator. For each node in the allocation, it SSH-launches a proxy (pmip), and each proxy in turn fork-execs the MPI ranks assigned to that node. 

The mpiexec–pmip control channel is always a TCP socket, regardless of the transport configured for MPI data traffic between ranks. Over this channel, the launcher and proxies communicate through Hydra commands. One example of such commands can be launch commands carrying the executable path, environment, and rank assignments. Hydra commands define types covering events such as process launch, PMI request forwarding, I/O relay, and exit status reporting.

The pmip-rank control channel is a Unix socketpair created by pmip during startup. Through PMI calls, pmip and proxies communicate over this PMI socket for rank setup and health management such as rank stdout and stderr redirection or business card exchanges (See \ref{sec:startup2}).

\subsection{Address Exchange}
\label{sec:startup2}

After the above architecture spawns, as part of communicator creation within MPI\_Init or MPI\_Session\_init calls, ranks discover one another through a key-value store coordinated by mpiexec to exchange business card, which is a string encoding a rank's transport endpoint, including its hostname, port, and connection tag. Each rank publishes its business card by issuing a PMI\_put on its socketpair. In Hydra's implementation, the proxy buffers these puts locally and flushes them to mpiexec as a single batch when the rank enters a PMI barrier. On the mpiexec side, it serves as the authoritative KVS store, ingesting each batch and tracks an epoch counter per contributing process. When every rank in the job has reached the same epoch, mpiexec signals all proxies, which unblock their ranks. Each rank then retrieves its peers' business cards via PMI\_get and opens peer connections for MPI data traffic. From this point on, ranks communicate directly with their peers over these transport connections, bypassing pmip and mpiexec entirely. This put–barrier–get cycle is the only point at which all ranks must globally converge through the process management layer.

\subsection{Teardown}
\label{sec:teardown}

Teardown reverses the startup sequence. When a rank calls MPI\_Finalize or MPI\_Session\_finalize, the MPI library closes all transport connections and releases their associated resources. After finalize returns, the rank retains no MPI network state; its only remaining tie to the runtime is the PMI socketpair. When the rank subsequently exits, pmip detects the termination and end-of-file on the socketpair and stdout/stderr pipes. Once all local ranks have terminated, pmip reports their exit statuses upstream via the command protocol and itself exits. 

\subsection{Observation}
\label{sec:hydra_observation}
MPI\_Session\_init / MPI\_Session\_finalize allows multiple rounds of communicator states creation and teardown (See \ref{sec:mpi_sessions}), this enables checkpointing opportunities for ranks between two MPI Sessions, in which they are ordinary user-space processes whose sole kernel-visible connection to the MPI runtime is a single Unix file descriptor. This is the window in which a process can be checkpointed without capturing any MPI transport state.
\section{Design}
\label{sec:design}        

\subsection{Workflow Overview}

\name enables selective migration of individual MPI ranks mid-execution without checkpointing the entire job. When a cloud provider issues a preemption notice for a node, \name migrates only the affected ranks to a replacement instance while the rest of the job pauses. The workflow has three stages: first, all ranks cooperatively quiesce, tearing down their MPI sessions and parking at a known-safe state (\ref{sec:cooperative_quiesce}); second, the process manager checkpoints the affected ranks, relocates them to a replacement proxy on the new node, and redirects the control plane; third, all ranks re-initialize their sessions and exchange updated endpoint addresses to establish fresh connections. Normal computation then resumes from where it left off (\ref{sec:migration_protocol}).

\name matches the granularity of the cloud preemption model. Per-instance migration prevents a preemption on a single instance from cascading into checkpointing and restarting every rank across every node. Only affected instances are checkpointed/restored, the remaining ranks experience only a brief quiesce stall rather than a full checkpoint and restart cycle, minimizing the downtime disruption.

\subsection{Cooperative Quiesce}
\label{sec:cooperative_quiesce}

\begin{algorithm}[t]
\caption{The \texttt{XMPI\_quiesce} procedure. During normal execution, the flag check at line 1 causes an immediate return. When migration is requested, the rank tears down its MPI session (lines 4–7), signals the proxy and blocks until relocation completes (lines 8–9), then reconstructs a fresh session (lines 11–13).}
\label{alg:quiesce}
\begin{algorithmic}[1]
\IF{migration\_requested $=$ 0}
    \RETURN \COMMENT{no-op during normal execution}
\ENDIF
\STATE MPI\_Barrier(comm)
\STATE MPI\_Comm\_free(comm)
\STATE MPI\_Group\_free(group)
\STATE MPI\_Session\_finalize(session)
\STATE write(pmi\_fd, \texttt{"xmpi\_quiesce\textbackslash n"})
\STATE read(pmi\_fd) \COMMENT{block until proxy responds}
\STATE migration\_requested $\leftarrow$ 0
\STATE MPI\_Session\_init(session)
\STATE MPI\_Group\_from\_session\_pset(session, group)
\STATE MPI\_Comm\_create\_from\_group(group, comm)
\end{algorithmic}
\end{algorithm}

To react to a preemption notice, \name exposes a CLI control utility, \texttt{hydra\_ctl}, for cloud admin to trigger a migration notification. At the time of being notified, all ranks must enter a quiesce state where they pause any computations and communications with no incomplete communications. This is due to MPI connections are inherently pairwise: if a migrating rank tears down its transport endpoint, the non-migrating peer holding the other end of that connection is left with a broken link. Rather than attempting to handle these half-open connections ( which would require error recovery logic in the MPI library), we require all ranks to tear down their sessions cooperatively, so that every connection is closed cleanly from both ends. 

\begin{algorithm}[t]
\caption{LULESH integration with \name. A single \texttt{XMPI\_quiesce} call at the iteration boundary (line 6) is the only application modification required.}
\label{alg:lulesh}
\begin{algorithmic}[1]
\STATE MPI\_init()
\STATE \textit{domain} $\leftarrow$ initialize simulation state
\WHILE{time $<$ stoptime}
    \STATE TimeIncrement(\textit{domain})
    \STATE LagrangeLeapFrog(\textit{domain}) \COMMENT{physics + MPI comms}
    \STATE \texttt{XMPI\_quiesce} \COMMENT{safe point: no in-flight messages}
\ENDWHILE
\STATE MPI\_finalize()
\end{algorithmic}
\end{algorithm}

We design an \texttt{XMPI\_quiesce} (Algorithm \ref{alg:quiesce}) interface for applications to insert calls at periodic safe points, marking where migration and its resulting quiesce may safely occur. Such safe points typically coincide with iteration boundaries in a simulation's main loop (Algorithm \ref{alg:lulesh}). The application developer, who understands the program's communication structure, identifies the safe quiesce points. When no migration notice is sent, \texttt{XMPI\_quiesce} functions as a no-op. All ranks will enter quiesce procedure when calling \texttt{XMPI\_quiesce} if there is notification received. Within the \texttt{XMPI\_quiesce}, ranks will call an MPI barrier to drain any potential in-flight messages or partially completed collectives. Migrating ranks utilizes MPI sessions API to gracefully cleanup old network context and start new sessions after migration, so that there is always no network states left at the time of checkpoint.

MPI Sessions are the enabling mechanism for \name's quiesce. Unlike \texttt{MPI\_Finalize} that permanently terminates a process's MPI participation, multiple session cycles can exist with a rank's lifetime (See \ref{sec:mpi_sessions}). This makes the quiesce-migrate-rejoin cycle or even multiple migration rounds possible. Because teardown and reconstruction go through the MPI implementation's own session path and \name never touches transport internals, \name is genuinely interconnect-agnostic.

We chose to require application participation by calling \texttt{XMPI\_quiesce} rather than interpose on the entire MPI API surface like MANA, because cooperative quiesce together with MPI's own session utilities makes clean communication state teardown and rebuild, eliminating the need for MPI state virtualization, replay logic, and transport-specific plugins, at the cost of a single application-inserted call.

\subsection{Migration Protocol}
\label{sec:migration_protocol}

\begin{figure*}[t]
\centering
\includegraphics[width=\textwidth]{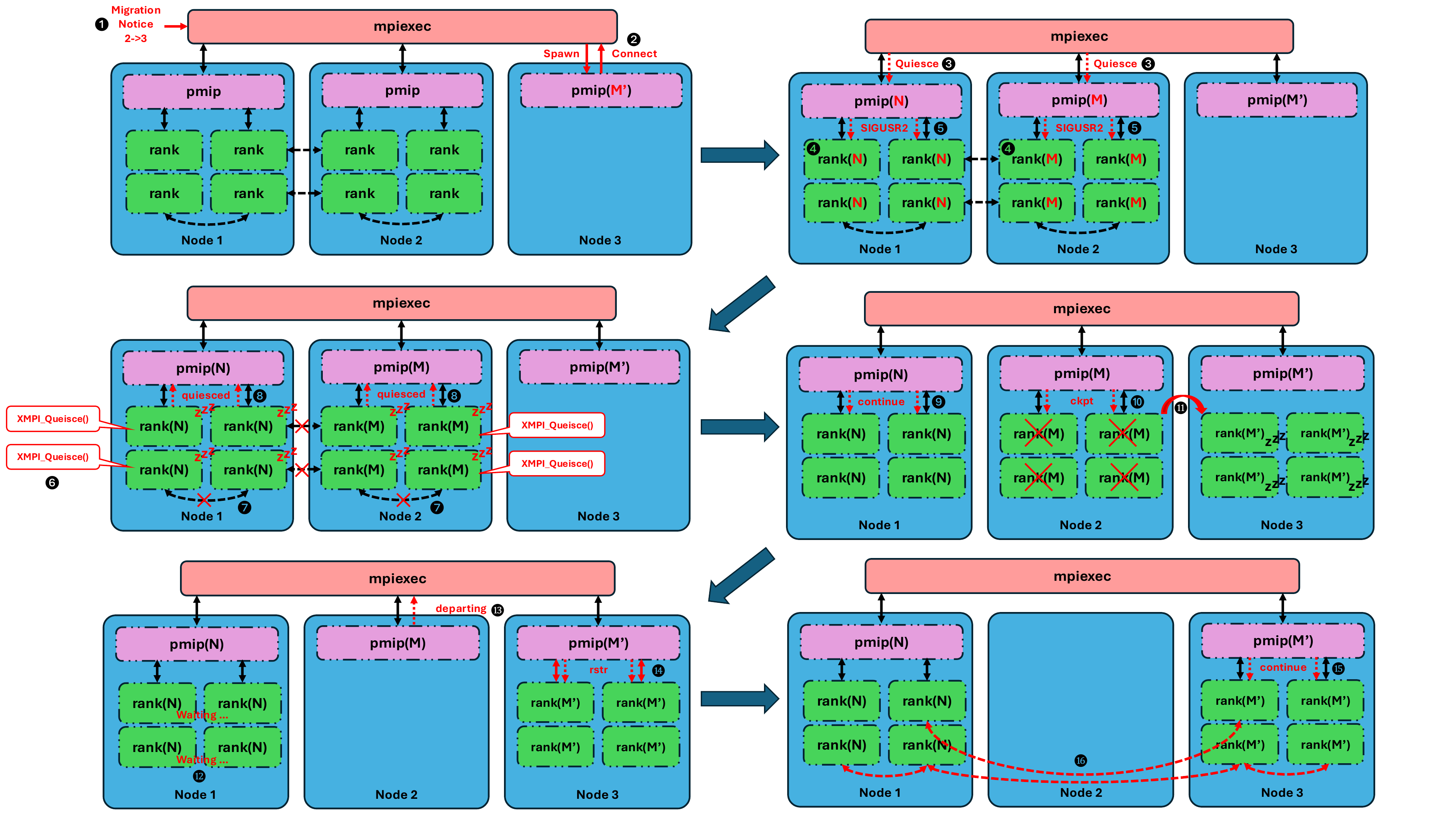}
\caption{Migration protocol timeline. }
\label{fig:migration_timeline}
\end{figure*}

We implement a new protocol on top of Hydra to conduct the selective migration procedure, shown in Figure \ref{fig:migration_timeline}. pmip(M) and rank(M) denote the proxy and ranks on the node being evacuated; pmip(N) and rank(N) denote proxies and ranks on nodes not involved in the migration; pmip(M') denotes the replacement proxy spawned on the target node; and rank(M') denotes the restored ranks after they have been relocated to the target node under pmip(M').

Migration begins when mpiexec receives a \texttt{hydra\_ctl} migration command that contains the migration target and destinations \circled{1}. It first spawns a replacement proxy, pmip(M'), on the target node. once pmip(M') connects \circled{2}, mpiexec initiates the quiesce sequence by sending out quiesce Hydra command \circled{3}. At receiving the command, migrating proxy pmip(M) and non-migrating proxy pmip(N) mark their ranks as migrating and non-migrating separately \circled{4}. Each proxy relays the notification to its ranks as a POSIX SIGUSR2 \circled{5}, notifying both rank(M) and rank(N) that a node preemption notification has received. The rank's handler sets a "migration\_requested" flag and returns immediately to continue executing application code with no checkpoint, no exit at this point.

At the next call site to \texttt{XMPI\_quiesce} \circled{6} (Section \ref{sec:cooperative_quiesce}), Both rank(M) and rank(N) first execute an MPI barrier to drain all in-flight messages, ensuring no peer's pending send is discarded by an early teardown. They then tear down their network states gracefully through MPI's own session API \circled{7}. Finally, all ranks write "xmpi\_quiesce" on their PMI socket to notify the proxies that they have reached a checkpoint-safe state \circled{8}. After this point, rank(M) and rank(N) behave differently. rank(M) block on a read from the PMI socket, waiting for a "xmpi\_continue" response of their pmip(M') after relocation completes. rank(M) remain parked this way while pmip(M) proceeds to checkpoint them. For rank(N), the proxy responds immediately with "xmpi\_continue" after rank(N) sends "xmpi\_quiesce", unblocking rank(N) to begin session re-initialization, they will wait at the PMI barrier embedded in the re-initialization calls until migrated ranks rank(M') arrive \circled{9} \circled{12}.     

After receiving "xmpi\_quiesce" from rank(M), pmip(M) invokes CRIU to checkpoint the processes \circled{10}. Because each rank has already finalized its MPI session before reaching this phase, the rank holds no MPI transport state at this point, only application-level memory and the PMI socket. CRIU writes rank images to disk together with the fd number of the rank's PMI socket (for later recovery purposes). CRIU then terminates the ranks and images are transferred by pmip(M) directly to the target node hosting pmip(M')\circled{11}. If multiple nodes are migrating, their pmip(M) initiate the transfers asynchronously and independently.

When pmip(M) has checkpointed all its local ranks, it sends a Hydra command upstream to mpiexec, distinguishing an intentional migration termination from crashes or exits \circled{13}. mpiexec then replaces its proxy table slot for pmip(M) to pmip(M'). From this point forward, every Hydra command that mpiexec sends to this proxy slot goes to pmip(M') on the target node.   
                                   
The replacement proxy pmip(M'), already running on the target node, receives a Hydra command from mpiexec and restores each rank from its CRIU image rather than fork-execing fresh processes \circled{14}. Since the restored rank's original PMI socket pointed to pmip(M), which no longer exists, pmip(M') creates fresh replacements and instructs CRIU to map them onto the fd numbers recorded during checkpoint. The restored rank wakes holding the same fd numbers it had before, but the other end of each now belongs to pmip(M') on the new node.

pmip(M') writes "xmpi\_continue" on the new PMI socket, unblocking the rank, which clears its migration flag and proceeds to \texttt{MPI\_Session\_init} \circled{15}. All ranks then re-execute the KVS cycle from Section \ref{sec:startup2}. All ranks republish business cards with rank(M') soliciting their new addresses on the target node. When the barrier releases, all ranks retrieve updated addresses and open fresh MPI connections \circled{16}. Execution resumes from the point where \texttt{XMPI\_quiesce} was called.

\subsection{Design Portability}
\label{sec:design_portability}

\begin{table}[t]
\centering
\caption{Process management abstractions across MPI implementations.}
\label{tab:pm_comparison}
\footnotesize
\begin{tabular}{llll}
\hline
 & \textbf{Proxy} & \textbf{PMI} & \textbf{Topology} \\
\hline
MPICH & pmip & PMI-1/2 & Flat \\
Open MPI & prted & PMIx & Tree (64) \\
Intel MPI & pmip & PMI-1/2 & Flat \\
Slurm & slurmstepd & PMI-1/2/x & Tree \\
Cray PALS & palsd & PMIx & Tree (32) \\
\hline
\end{tabular}
\end{table}

\name's migration protocol depends on three abstractions seen in Hydra: a central coordinator that holds process management state, per-node proxies that relay traffic between the coordinator and ranks, and a PMI channel connecting each rank to its local proxy. These abstractions are not Hydra-specific, they are commonly present in major MPI process managers (Table \ref{tab:pm_comparison}). Open MPI's PRRTE uses prted as its per-node proxy with PMIx over Unix sockets, Slurm's slurmstepd serves the same role with configurable PMI-1, PMI-2, or PMIx support, and Intel MPI uses an unmodified fork of Hydra. They share similar structures, and thus \name's design is highly portable to other process management implementation. The rank-side code uses only standard MPI Sessions API calls and requires no modification inside implementations, thus \name can work on top of any standard-conformant MPI.

The one architectural difference worth addressing is proxy topology. Hydra connects all proxies directly to mpiexec in a flat tree, while PRRTE organizes proxies into a radix tree with a default fanout of 64, and Cray PALS uses a tree with fanout 32. Under a hierarchical topology, the coordinator-to-proxy command path would traverse intermediate proxies, requiring them to forward migration commands rather than receive them directly. However, this distinction is largely academic for real workloads: facility data from NERSC and NREL \cite{patki2025nersc, rodrigo2018jobsizes} shows that 95–98\% of HPC jobs by count run on fewer than 64 nodes, with medians of 1–4 nodes. At these scales, PRRTE's tree collapses to a single level, which is functionally identical to Hydra's flat topology. Extending XMPI to multi-level trees is straightforward but orthogonal to our core contribution, and we defer it to future work (Section \ref{sec:future_work}).

\section{Evaluation}

This section seeks to answer the following questions:
\begin{itemize}
\item \textbf{Q1:} What overhead does \name introduce during normal execution when no migration occurs?
\item \textbf{Q2:} How long are migrating ranks unavailable during a migration event, and what dominates the downtime?
\item \textbf{Q3:} How does the stall imposed on non-migrating ranks scale with per-node rank count, job size, and evacuation width?
\item \textbf{Q4:} Does \name remain stable across repeated migrations, with no state leaks or degradation?
\end{itemize}

\subsection{Experimental Setup}
\label{sec:experimental_setup}

\textbf{Hardware.}
We use a 9-node cluster of identical machines, each equipped with two Intel Xeon Gold 6242 CPUs (16 cores per socket, 2.80\,GHz base, 3.90\,GHz turbo), 192\,GB DDR4 RAM, and 10\,Gbps Ethernet. Nodes run Ubuntu 22.04 (kernel 5.15.0).

\textbf{Software.}
MPICH 4.2.2 with the CH4:OFI communication device (\texttt{FI\_PROVIDER=tcp}). CRIU version 4.2. All benchmarks are compiled with \texttt{-O3}. Image transfer is handled by \texttt{scp}; reducing transfer latency is orthogonal to \name's design and left to the underlying transport. 

\textbf{Topology.}
All experiments run on a nine-node cluster: one node is dedicated to
the launcher and runs \texttt{mpiexec} only, hosting no compute ranks,
while the remaining eight nodes serve as compute nodes and migration
targets. We denote by $C$ the number of compute nodes, each running a
single Hydra proxy (\texttt{pmip}) that hosts a fixed number of ranks
per node ($\mathit{rpn}$), and by $M$ the number of proxies evacuated in
a migration event; the two are bounded by $C + M \le 8$ and $M \le C$.
Because Hydra runs one proxy per node, each evacuated proxy is relocated
to a distinct idle target, so $M$ concurrent evacuations consume $M$ free
target nodes. Source and target nodes are kernel-matched so that CRIU can
restore each checkpoint image on its target. During a migration, each
source proxy transfers its rank images directly to its paired target over
\texttt{scp}, independently of \texttt{mpiexec} and of every other
migrating proxy; the $M$ transfers therefore proceed concurrently rather
than being relayed serially through the launcher host.

\textbf{Benchmarks.}
We evaluate four proxy applications from the Mantevo~\cite{heroux2009mantevo} and ECP suites, selected to span distinct computational and communication patterns (Table~\ref{tab:benchmarks}):
\begin{itemize}
\item \textbf{LULESH 2.0}~\cite{karlin2013lulesh}: a shock hydrodynamics code with structured halo exchange between neighboring subdomains.
\item \textbf{CoMD 1.1}~\cite{comd}: a classical molecular dynamics code with irregular, data-dependent neighbor communication as atoms migrate between spatial domains.
\item \textbf{HPCCG 1.0}~\cite{heroux2007hpccg}: a conjugate gradient solver dominated by sparse matrix-vector products and global allreduce operations.
\item \textbf{miniAMR 1.0}~\cite{sasidharan2016miniamr}: an adaptive mesh refinement code with dynamic neighbor communication and communicator splitting as blocks are refined and coarsened.
\end{itemize}
All four were ported to \name by replacing \texttt{MPI\_Init}/\texttt{MPI\_COMM\_WORLD} with MPI Sessions and inserting a single \texttt{XMPI\_quiesce} call at each application's main iteration boundary. We use scp as the image transferring tools since \name do not focus on optmizing image transferring latency.

\begin{table}[t]
\centering
\caption{Benchmark applications and their communication characteristics.}
\label{tab:benchmarks}
\footnotesize
\begin{tabular}{llll}
\hline
\textbf{Benchmark} & \textbf{Domain} & \textbf{Comm. Pattern}\\
\hline
LULESH 2.0 & Shock hydro. & Structured halo exchange \\
CoMD 1.1 & Molecular dyn. & Irregular neighbor exch. \\
HPCCG 1.0 & CG solver & SpMV + allreduce \\
miniAMR 1.0 & Adaptive mesh & Dynamic comm + split\\
\hline
\end{tabular}
\end{table}

\subsection{Normal-Run Overhead (Q1)}
\label{sec:eval_overhead}

To measure the overhead \name introduces during normal execution, we compare the \name-instrumented binary (MPI Sessions, signal handler, and \texttt{XMPI\_quiesce} called once per timestep) against an unmodified baseline using \texttt{MPI\_Init}/\texttt{MPI\_COMM\_WORLD}. Both binaries are linked against the same MPICH build. We run each benchmark on a single node with 8 ranks.
\begin{figure}[t]
\centering
\includegraphics[width=\columnwidth]{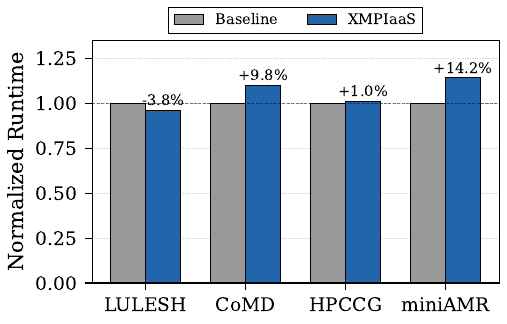}
\caption{Normalized runtime of \name versus baseline (no migration triggered). Differences range from $-3.8\%$ to $+14.2\%$ across benchmarks, consistent with instruction cache alignment noise rather than systematic overhead.}
\label{fig:overhead}
\end{figure}

Figure~\ref{fig:overhead} shows the normalized runtime for each benchmark. The direction of the difference flips between benchmarks: LULESH runs 3.8\% \emph{faster} under \name, CoMD is 9.8\% slower, HPCCG is within 1.0\%, and miniAMR is 14.2\% slower. This pattern where the sign of the difference varies across binaries despite identical hardware is characteristic of instruction cache alignment noise rather than real overhead. The per-iteration cost of \texttt{XMPI\_quiesce} when no migration is pending consists of a single volatile read of the \texttt{migration\_requested} flag followed by an untaken branch, which is unmeasurable at the granularity of our benchmarks. We conclude that \name introduces no meaningful overhead during normal execution.

\subsection{Migration Downtime Breakdown (Q2)}
\label{sec:eval_breakdown}

To understand what dominates migration downtime, we instrument the migration path with millisecond-resolution timestamps (\texttt{clock\_gettime}) and decompose the total downtime of migrating ranks into four phases: quiesce (barrier + session teardown), CRIU dump, image transfer (\texttt{scp}), and restore + reconnect (CRIU restore, \texttt{xmpi\_continue}, session re-initialization, and KVS address exchange). We migrate 4 ranks from one node to another across all four benchmarks. Migration downtime is the interval from the first CRIU dump to the last restore-and-reconnect across the migrating ranks.

\begin{figure}[t]
\centering
\includegraphics[width=\columnwidth]{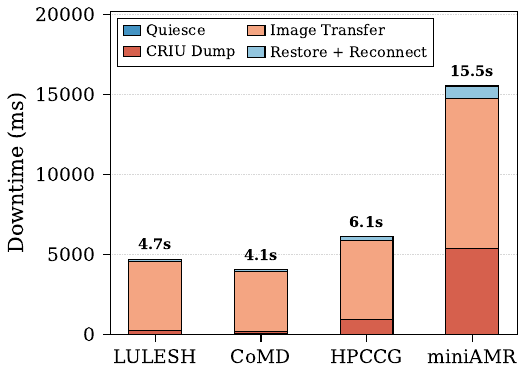}
\caption{Migration downtime breakdown by phase. Image transfer via \texttt{scp} dominates in all cases (60--92\%). Quiesce is negligible ($<$55\,ms). CRIU dump time scales with per-rank memory footprint.}
\label{fig:breakdown}
\end{figure}

Figure~\ref{fig:breakdown} shows the stacked breakdown. Three findings stand out. First, image transfer dominates total downtime across all benchmarks, accounting for 91\% in LULESH, 92\% in CoMD, 81\% in HPCCG, and 60\% in miniAMR. This is an artifact of our testbed's 10\,Gbps Ethernet and the use of \texttt{scp}; on systems with a shared parallel filesystem, this phase is eliminated entirely since both the source and target nodes can access the checkpoint images directly. Second, the quiesce phase is negligible in all cases: 17\,ms for LULESH, 55\,ms for CoMD, 9\,ms for HPCCG, and 7\,ms for miniAMR. This confirms that cooperative session teardown via the MPI Sessions API adds minimal latency. Third, CRIU dump time scales with per-rank memory footprint: miniAMR, which maintains the largest working set due to its refined mesh blocks, incurs 5.4\,s of dump time versus 260\,ms for LULESH.

Excluding image transfer, the core migration machinery (quiesce + dump + restore) completes in 424\,ms for LULESH, 316\,ms for CoMD, 1.2\,s for HPCCG, and 6.2\,s for miniAMR. These times represent the irreducible cost of the migration protocol itself and would be the total downtime on a system with shared storage.

\subsection{Non-Migrating Rank Stall (Q3)}
\label{sec:eval_stall}

Section~\ref{sec:eval_breakdown} established that migration downtime is
dominated by image transfer and that CRIU dump scales with per-rank
memory footprint. This subsection characterizes how the resulting
downtime scales along three axes: the number of ranks hosted on a
migrating node ($\mathit{rpn}$), the number of compute nodes in the job
($C$), and the number of proxies evacuated concurrently ($M$). The
central finding is that downtime is governed almost entirely by
$\mathit{rpn}$ (the rank count on the migrating node) and is
insensitive to both total job size and evacuation width.

\begin{table}[t]
\centering
\caption{Flagship scaling configurations (CoMD, HPCCG). Each set varies one
axis (bold) through the shared center $C{=}4$, $\mathit{rpn}{=}16$, $M{=}1$.
Sets~B and~C are the $M{=}1$ row and $\mathit{rpn}{=}16$ column of the Set~D grid.}
\label{tab:configs_flagship}
\footnotesize
\begin{tabular}{lllll}
\hline
\textbf{Set} & \textbf{Nodes ($C$)} & \textbf{Ranks/Node} & \textbf{$M$} & \textbf{Total Ranks} \\
\hline
A & \textbf{2--7} & 16 & 1 & 32--112 \\
B & 4 & \textbf{2--32} & 1 & 8--128 \\
C & 4 & 16 & \textbf{1--4} & 64 \\
D & 4 & 4--32 & \textbf{1--4} & 16--128 \\
\hline
\end{tabular}
\end{table}

\begin{table}[t]
\centering
\caption{LULESH scaling configurations. The perfect-cube rank constraint
replaces the fixed-$\mathit{rpn}$ node sweep with strong- and weak-scaling
variants; swept axis in bold.}
\label{tab:configs_lulesh}
\footnotesize
\begin{tabular}{lllll}
\hline
\textbf{Set} & \textbf{Nodes ($C$)} & \textbf{Ranks/Node} & \textbf{$M$} & \textbf{Total Ranks} \\
\hline
A$'$ (strong) & \textbf{1--4} & 2--64 & 1 & 8, 27, 64 \\
B$'$ (weak)   & \textbf{2--4} & 4--16 & 1 & 8--64 \\
C (width)     & 3--4 & 2--16 & \textbf{1--4} & 8, 27, 64 \\
\hline
\end{tabular}
\end{table}

We sweep three axes — per-node rank count ($\mathit{rpn}$),
compute-node count ($C$), and evacuation width ($M$) — through a shared
center ($C{=}4$, $\mathit{rpn}{=}16$, $M{=}1$); Table~\ref{tab:configs_flagship}
lists the configurations for the flexible-rank benchmarks. LULESH's
perfect-cube rank constraint ($\text{total}\in\{8,27,64\}$) precludes a
fixed-$\mathit{rpn}$ node sweep, so we scale its job by strong and weak
scaling instead (Table~\ref{tab:configs_lulesh}).

\begin{figure}[t]
\centering
\includegraphics[width=0.8\columnwidth]{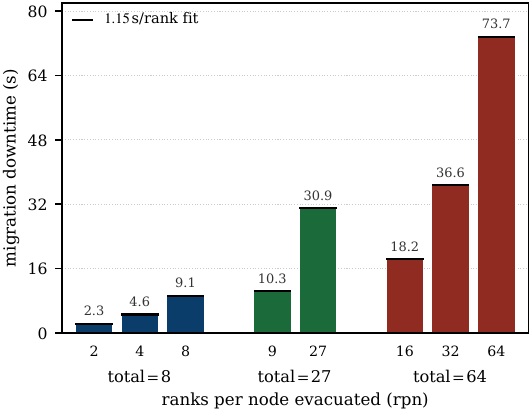}
\\[3pt]
\includegraphics[width=0.8\columnwidth]{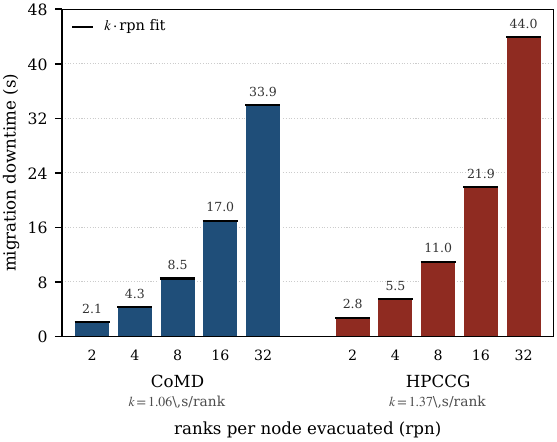}
\caption{Migration downtime versus per-node rank count ($\mathit{rpn}$):
LULESH (top); CoMD and HPCCG (bottom).}
\label{fig:rpn_bars}
\end{figure}

Figure~\ref{fig:rpn_bars} plots migration downtime against per-node rank
count. In every benchmark downtime is linear in $\mathit{rpn}$ and passes
through the origin: the fitted $k\cdot\mathit{rpn}$ line (black caps) matches
the measured bars, and the intercept is within noise of zero (e.g.\
$+0.004$\,s for CoMD), indicating that the one term expected to grow with
total job size, the global KVS address re-exchange, is negligible at these
scales. The slope $k$ is benchmark-specific, ranging from $1.06$\,s/rank for
CoMD to $1.37$\,s/rank for HPCCG, and tracks per-rank image size: HPCCG's
sparse $64^3$ blocks yield the largest checkpoints and the steepest slope,
CoMD's compact atom lists the smallest and shallowest. This is the direct
consequence of the transfer-bound behavior established in
Section~\ref{sec:eval_breakdown} — at a fixed per-rank footprint, the bytes
evacuated from a node scale with its rank count, and transfer time with them.
LULESH (top), whose perfect-cube constraint ties rank count to job size,
makes the same point across three job sizes at once: bars spanning totals of
$8$, $27$, and $64$ ranks all fall on a single $1.15$\,s/rank line, so a
given per-node density incurs the same downtime regardless of the total job
it belongs to. Downtime is thus governed by the rank count on the migrating
node alone, not the size of the job around it.

\begin{figure*}[t]
\centering
\includegraphics[width=0.32\textwidth]{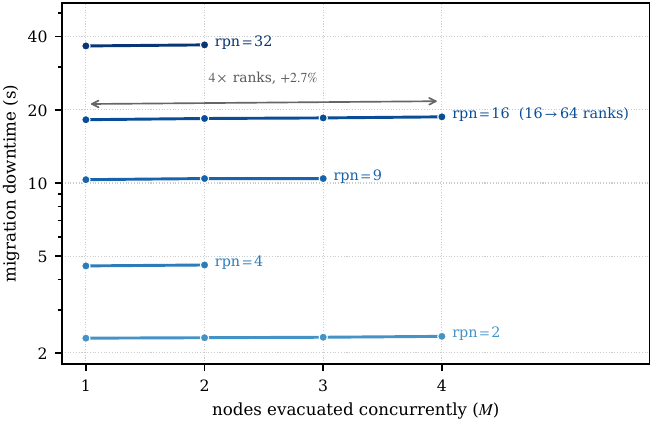}\hfill
\includegraphics[width=0.32\textwidth]{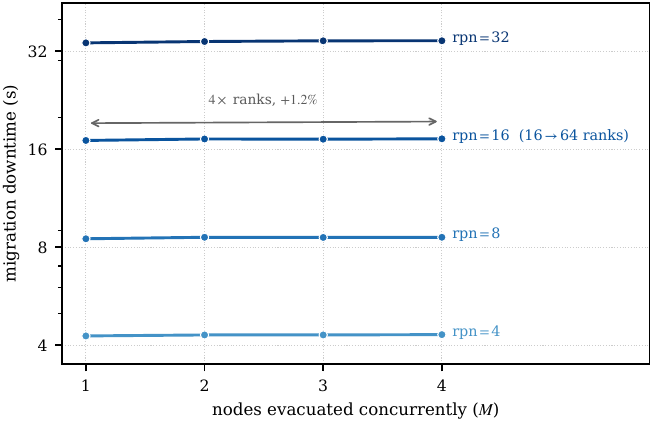}\hfill
\includegraphics[width=0.32\textwidth]{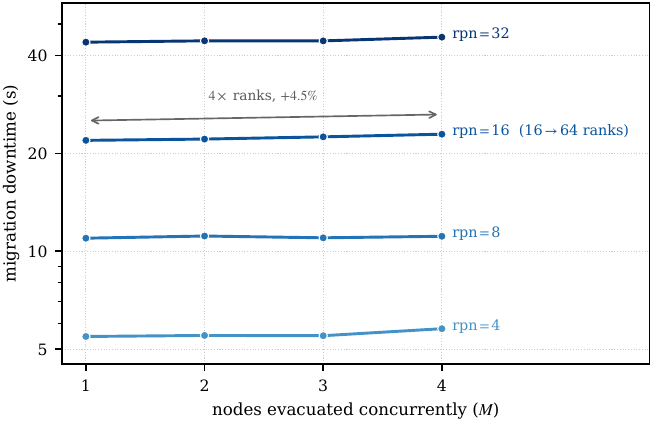}
\caption{Migration downtime versus evacuation width $M$: LULESH (left),
CoMD (center), HPCCG (right).}
\label{fig:M_scaling}
\end{figure*}

Figure~\ref{fig:M_scaling} varies the evacuation width $M$ — the number of
proxies relocated concurrently — from one to four, holding per-node density
fixed. At every $\mathit{rpn}$ level and in all three benchmarks, downtime
stays essentially flat as $M$ grows: evacuating four nodes at once costs
almost the same as evacuating one, even though four times as many ranks are
in flight. At $\mathit{rpn}{=}16$, going from $M{=}1$ to $M{=}4$ quadruples
the ranks relocated ($16\to64$) while downtime rises only $+2.7\%$ for
LULESH, $+1.2\%$ for CoMD, and $+4.5\%$ for HPCCG. This flatness follows
directly from the migration protocol: each source proxy transfers its
checkpoint images directly to its paired target, independently of
\texttt{mpiexec} and of every other migrating proxy
(Section~\ref{sec:migration_protocol}), so the $M$ transfers proceed in
parallel and total downtime is the maximum over them rather than their sum.
The small residual increase is ordered by per-rank image size — largest for
HPCCG, smallest for CoMD — the signature of mild contention on the shared
network fabric as concurrent image volume rises; it remains under $5\%$ even
in the most demanding case (HPCCG at $\mathit{rpn}{=}32$, $M{=}4$: 128 ranks
and $\sim$5.4\,GB pushed across four concurrent links). \name thus lets an
operator evacuate an arbitrary number of simultaneously preempted nodes for
roughly the cost of evacuating one.

\begin{figure*}[t]
\centering
\includegraphics[width=0.32\textwidth]{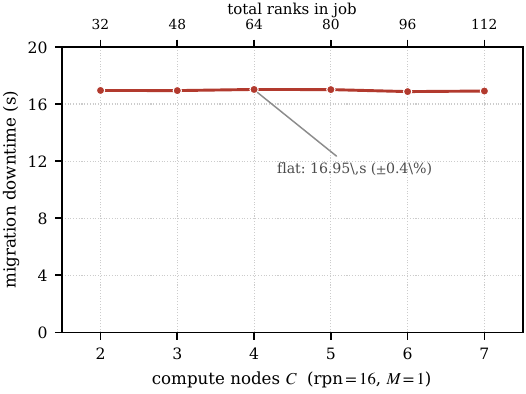}\hspace{0.04\textwidth}%
\includegraphics[width=0.32\textwidth]{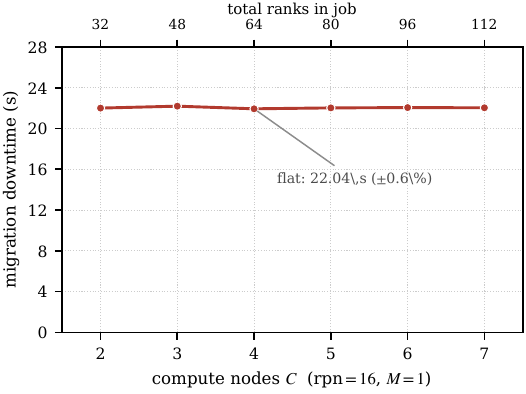}
\caption{Migration downtime versus compute-node count $C$ at fixed
$\mathit{rpn}{=}16$, $M{=}1$: CoMD (left), HPCCG (right).}
\label{fig:C_scaling}
\end{figure*}

Figure~\ref{fig:C_scaling} isolates the effect of job size by holding
per-node density fixed ($\mathit{rpn}{=}16$, $M{=}1$) and scaling the job
from two to seven compute nodes ($32$ to $112$ total ranks). Downtime stays
flat throughout, at $16.95$\,s ($\pm0.4\%$) for CoMD and $22.04$\,s
($\pm0.6\%$) for HPCCG. Since only one node is evacuated, the cost is set by
the $16$ ranks it holds and is indifferent to how many other nodes
participate. A preemption on one instance therefore never scales into a
job-wide cost, which is the essence of selective migration.

\subsection{Repeated Migration Stability (Q4)}
\label{sec:eval_stability}

To verify that \name introduces no state leaks or performance degradation over multiple migration cycles, we run LULESH (8 ranks, 2 nodes, nx=45) and trigger 8 consecutive migrations at 30-second intervals, alternating the migrating proxy between two target nodes. We record the downtime of each migration event and the resident set size (VmRSS) of the mpiexec coordinator process.

\begin{figure}[t]
\centering
\includegraphics[width=\columnwidth]{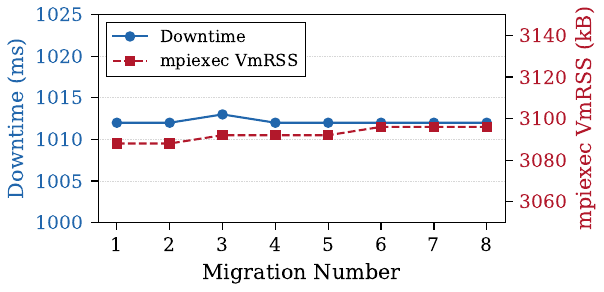}
\caption{Downtime and mpiexec memory footprint over 8 consecutive migrations. Downtime is flat at 1012\,ms ($\sigma < 1$\,ms). Memory grows by only 8\,kB total, confirming no state leak.}
\label{fig:stability}
\end{figure}

Figure~\ref{fig:stability} shows that migration downtime is effectively constant at 1012\,ms with sub-millisecond variance ($\sigma < 1$\,ms) across all 8 events. The mpiexec VmRSS increases by only 8\,kB over the entire sequence, confirming that the proxy replacement and epoch reclamation logic in mpiexec does not leak state. The fd-swap handover replaces the old proxy's control socket with the new one in place, and the KVS epoch entries are reclaimed rather than accumulated. These results demonstrate that \name can sustain repeated migrations indefinitely without degradation, as would be required in a long-running cloud deployment subject to periodic preemption events.

\section{Future Work}
\label{sec:future_work}

\name currently requires the application developer to insert \texttt{XMPI\_quiesce} calls at safe points, which is a reasonably trivial burden for iterative scientific codes, but a barrier for complex applications with irregular communication patterns or deeply nested call hierarchies where safe points are not immediately obvious. An automatic approach that identifies quiescent program points through static compiler analysis such as detecting loop boundaries where no MPI request is outstanding, or through lightweight runtime profiling that tracks in-flight message counts, would broaden applicability to unmodified legacy codes and remove the programming model requirement entirely.

Our implementation targets Hydra's flat proxy topology, in which all proxies connect directly to mpiexec. Extending the migration protocol to hierarchical proxy trees ( as used by PRRTE (fanout 64) and Cray PALS (fanout 32) at scale ) would require intermediate proxies to forward migration commands rather than receive them directly, and the fd-swap handover would need to propagate through the tree rather than occur at a single level. This extension would enable \name to operate on jobs spanning hundreds of nodes, though as discussed in Section~\ref{sec:design_portability}, fewer than 5\% of HPC jobs by count exceed the threshold where hierarchical routing engages.

The serverless deployment model described in Section~\ref{sec:cloud} is a natural extension of our checkpoint/restore machinery. Each MPI rank would run as an independent function invocation, with \texttt{XMPI\_quiesce} triggering a checkpoint to shared storage as the function's time limit approaches. Fresh invocations would restore from the images and resume computation, making execution appear continuous while conforming to the serverless platform's stateless, time-bounded execution model. The key challenges are adapting the process manager to operate across ephemeral function instances and keeping checkpoint/restore overhead small relative to the function's available runtime.
\section{Related Work}
\label{sec:related_work}
 
\textbf{System-level checkpoint/restart.} 
BLCR ~\cite{hargrove2006blcr} provided kernel-level checkpoint/restart for Linux clusters and was integrated with several MPI implementations including LAM/MPI ~\cite{squyres2003lam} and Open MPI, but required kernel module maintenance and was eventually abandoned as kernel APIs evolved. CRIU~\cite{criu} succeeded BLCR by operating primarily in userspace, but neither tool can handle MPI's network states, InfiniBand queue pairs, RDMA memory registrations, and device-backed memory mappings are all opaque to their checkpoint machinery. DMTCP~\cite{ansel2009dmtcp} extended transparent checkpointing to distributed and multithreaded applications with a coordinator-based architecture, and serves as the foundation for MANA's split-process approach.

\noindent\textbf{Whole-job checkpoint/restart.} MANA \cite{garg2019mana} transparently checkpoints/restores MPI applications by interposing on the entire MPI API interface, maintaining a virtualized bookkeeping of internal MPI states. It employs a split-process architecture that separates the application from the MPI library through a proxy layer, discarding the network contexts that are difficult to preserve transparently at checkpointing time. MANA~2.0~\cite{xu2023mana2} improved scalability by introducing collective vector clocks to determine safe synchronization points algorithmically. MANA's intrusive approach comes at considerable deployment complexity that it must track and replay every MPI object's lifecycle, requires a dedicated DMTCP coordinator daemon \cite{ansel2009dmtcp} alongside the job, and each new transport (TCP, InfiniBand, Slingshot) demands a separate network-virtualization plugin. 
Application-level checkpointing libraries such as SCR \cite{moody2010scr} and FTI \cite{bautista2011fti} enable efficient state serialization to local or parallel storage, but require the programmer to manually identify and save all relevant state.
Works such as MANA, SCR and FTI all treat recovery as a whole-job event and necessitate a full job restart from the last checkpoint. A preemption on a single instance cascades into a full teardown.

\noindent\textbf{Malleability and resource reconfiguration.} Malleable MPI frameworks allow a running job to change its process count in response to a resource manager's decisions. Elastic MPI \cite{compres2016elastic} extends MPICH and Slurm so that applications periodically poll for reconfiguration events, admitting or releasing processes as the scheduler dictates. DMR \cite{iserte2018dmr, iserte2021dmrlib} builds on \texttt{MPI\_Comm\_spawn} to expand or shrink a job at application-declared reconfiguration points, with a companion library redistributing application data into the new decomposition. Dynamic PSets \cite{huber2024dpp} uses a dynamic resource manager to grant or reclaim an allocation and directs the job's runtime daemon to spawn processes into it. All three treat the departing rank's state as discarded, and the application must both run correctly at the new process count and redistribute its data accordingly. Charm++ ~\cite{kale1993charm}\cite{kale1996charm} and its MPI interface AMPI \cite{huang2003ampi} overdecomposes applications into virtual ranks, each rank is a serializable object that the load balancer can relocate between nodes \cite{chakravorty2006proactive}. Recent work has extended Charm++ to handle spot instance preemption on cloud platforms~\cite{bhosale2025charm} and elastic job scheduling~\cite{bhosale2025elastic}. However, AMPI still requires globals privatization and an over-decomposed launch, migrating its own virtual ranks rather than an unmodified MPI rank.

\noindent\textbf{Contained Recovery.} 
User-Level Fault Mitigation (ULFM)~\cite{bland2013ulfm, losada2020ulfm} extends the MPI standard with primitives for detecting process failures and repairing communicators, allowing applications to continue after a rank crashes. However, ULFM provides only the detection and communicator-repair mechanism, it preserves no process state and requires applications to implement their own recovery logic. SPBC \cite{ropars2013spbc} combines coordinated checkpointing with message logging, so that a failure rolls back only the affected process group, which replays its logged messages while the rest of the job waits in place rather than restarting. The logging overhead is paid on every message of every run, whether or not a failure ever occurs, and the recovered ranks must still re-execute the work performed since their last checkpoint.

\noindent\textbf{Process-level migration.} 
Wang et al. \cite{wang2008proactive} migrate the ranks of a single deteriorating node to a spare while the remaining ranks quiesce and resume in place, using BLCR-based \cite{hargrove2006blcr} memory precopy and an MPI-level drain of in-flight messages. However, their design presumes a long-deprecated MPI runtime architecture, requesting the destination to already host a daemon in the job's control plane. Wang et al. \cite{wang2025livemigration} migrate a containerized MPI rank group with CRIU. However, their work makes no mention to how MPI runtime handles the migration, and their evaluation only migrates single container MPI job.

\section{Conclusion}
\label{sec:conclusion}

We present \name, a novel cooperative migration framework that
enables selective relocation of MPI ranks mid-execution without
checkpointing the entire job. By combining application-identified
quiesce points with MPI's own Sessions API for transport teardown,
\name reduces a migrating rank to an ordinary userspace process,
trivially checkpointable by CRIU without MPI-specific plugins or
state virtualization. On the process manager side, a lightweight
protocol of three new Hydra commands and two out-of-band sentinel
strings coordinates the full migration lifecycle: quiescence,
selective checkpointing, proxy replacement via a single fd swap, and
seamless reconnection through the existing KVS address exchange.

Our evaluation across four proxy applications demonstrates that
\name's cooperative design achieves its goals. The quiesce phase adds
less than 55\,ms of latency. The core migration machinery completes in 0.3--6.2\,s depending on
per-rank memory footprint. Migration downtime depends only on the migrating node's rank count, and non-migrating ranks are never checkpointed. The system sustains
repeated migrations with constant 1012\,ms downtime and no state
leaks over eight consecutive cycles. Normal-execution overhead is
indistinguishable from noise.

Two properties make \name practical for cloud deployment. First, its
per-instance granularity matches the cloud preemption model: only the
affected node's ranks are checkpointed, while the rest of the job
continues with minimal disruption. Second, its reliance on standard
MPI Sessions API calls on the rank side and common process management
abstractions on the runtime side makes the design portable across MPI
implementations. As cloud platforms increasingly host HPC workloads on
preemptible infrastructure, \name provides a path toward resilient,
cost-effective MPI execution without sacrificing the programming model
that three decades of scientific software depend on.

\bibliographystyle{ACM-Reference-Format}
\bibliography{ref}

\end{document}